\documentclass{jfm}
\usepackage{natbib}
\usepackage{graphicx}
\title[Bed Erosion by Laminar Flow]{Erosion of a granular bed driven \\
  by laminar fluid flow}

\author[A. E. Lobkovsky, A. V. Orpe, R. Molloy, A. Kudrolli and D. H.
Rothman]{A\ls L\ls E\ls X\ls A\ls N\ls D\ls E\ls R\ns E.\ns L\ls O\ls B\ls
  K\ls O\ls V\ls S\ls K\ls Y$^1$, A\ls S\ls H\ls I\ls S\ls H\ns V.\ns
  O\ls R\ls P\ls E$^2$,\\ R\ls Y\ls A\ls N\ns M\ls O\ls
  L\ls L\ls O\ls Y$^2$,\ns A\ls R\ls S\ls H\ls A\ls D\ns K\ls U\ls
  D\ls R\ls O\ls L\ls L\ls I$^2$ \and D\ls A\ls N\ls I\ls E\ls L\ns
  H.\ns R\ls O\ls T\ls H\ls M\ls A\ls N$^1$}

\affiliation{$^1$Department of Earth, Atmospheric, and Planetary
  Sciences, Massachusetts Institute of Technology, Cambridge, MA 02139
  \\[\affilskip] $^2$Physics Department, Clark University, Worcester,
  MA 01610}

\date{\today}

\begin{document}

\maketitle

\begin{abstract}
  Motivated by examples of erosive incision of channels in sand, we
  investigate the motion of individual grains in a granular bed driven
  by a laminar fluid to give us new insights into the relationship
  between hydrodynamic stress and surface granular flow.  A closed
  cell of rectangular cross-section is partially filled with glass
  beads and a constant fluid flux $Q$ flows through the cell.  The
  refractive indices of the fluid and the glass beads are matched and
  the cell is illuminated with a laser sheet, allowing us to image
  individual beads.  The bed erodes to a rest height $h_r$ which
  depends on $Q$.  The Shields threshold criterion assumes that the
  non-dimensional ratio $\theta$ of the viscous stress on the bed to
  the hydrostatic pressure difference across a grain is sufficient to
  predict the granular flux.  Furthermore, the Shields criterion
  states that the granular flux is non-zero only for $\theta >
  \theta_c$.  We find that the Shields criterion describes the
  observed relationship $h_r \propto Q^{1/2}$ when the bed height is
  offset by approximately half a grain diameter.  Introducing this
  offset in the estimation of $\theta$ yields a collapse of the
  measured Einstein number $q^*$ to a power-law function of $\theta -
  \theta_c$ with exponent $1.75 \pm 0.25$.  The dynamics of the bed
  height relaxation are well described by the power law relationship
  between the granular flux and the bed stress.
\end{abstract}

\section{Introduction}
\label{sec:introduction}

The response of a granular bed to forcing by a fluid which flows
through and over the bed has been the subject of continuous inquiry
for over a century (see, for example
\cite{graf71:_hydraulics,yalin77:_mechanics}).  This phenomenon is at
the centre of a wide range of practical and fundamental problems.
Predicting the granular flux for a known fluid flow is important in
understanding how beaches \citep{bailard81:_beach,komar98:_beach},
rivers \citep{murray94:_braided} and deltas
\citep{kenyon85:_delta_prograding,parker98:_fans} evolve, mountains
erode \citep{burbank96:_bedrock} and landscapes form
\citep{howard94:_detachment}.  Sedimentary records cannot be
deciphered without a working understanding of the combined
fluid/granular (two-phase) flow \citep{blum00:_stratigraphy}.  The
microscopic details of the bed's response to forcing by a fluid raise
important fundamental questions about the nature of fluid flow near a
rough \citep{grass71:_turb_flow,jimenez04:_flow_rough} and/or
permeable \citep{brinkman49,beavers67:_bc} wall, the motion of a grain
on a rough surface \citep{samson99:_bumpy,quartier00:_grain_sandpile},
and the dynamics of granular avalanches
\citep{douady02:_surface_flows,bonamy02:_surf_flow}.

Given some measure of the fluid flow intensity, one would like to
predict the granular flux from the properties of the granular material
such as size, shape, friction coefficient, bed packing, etc.  An
important aspect of the problem to consider first is the onset of
granular flow.  Just as the surface of a granular pile driven by
gravity alone relaxes to the angle of repose
\citep{jaeger89:_relax_angle_repos}, granular beds driven by fluid
flow are thought to be static below a certain threshold fluid flux.
Several empirical curves relating some measure of the fluid forcing at
the onset of granular flow to the grain properties have been proposed
\citep{shields36,vanoni64:_meas,yalin77:_mechanics}.  Numerous
experimental studies aimed at computing these curves (summarised in
\cite{miller77:_shields_review} or \cite{buffington97:_systematic},
for example) are difficult to interpret since measured critical
properties (such as the Shields parameter) vary by as much as a factor
of three from study to study.  The disparate and subjective
definitions of the onset of granular flow used by the researchers are
frequently used to explain the observed scatter
\citep{shvidchenko00:_flume,paphitis01:_threshold}.

Even the existence of the threshold fluid forcing below which the
granular flux is identically zero is a subject of a lively ongoing
debate \citep{graf77:_small_flux,rijn89:_delft}.  When the flow is
turbulent, the local bed forcing is stochastic
\citep{papanicolaou01:_turb_rough} and a strong case, supported by
data, can be made for the presence of a granular flux (albeit
vanishingly small) for any mean value of the fluid forcing
\citep{rijn89:_delft}.  The omnipresence of a granular flux poses a
two-fold problem.  First, an arbitrary threshold grain flux
\citep{neill69:_quant}, dislodging rate \citep{shvidchenko00:_flume},
or dislodging probability \citep{dancey02:_probab} must be introduced
to characterise the onset.  Whereas the study of
\cite{graf77:_small_flux} suggests that the granular flux is
exponentially small below some threshold driving, no systematic
attempt, to our knowledge, has been made to quantify the transition in
the functional dependence of the granular flux on driving.  Second,
the ever-present granular flux results in the evolution of the surface
packing.  This ``ageing'' or ``armouring'' of the granular bed leads
to a decrease in the granular flux and an increase in the perceived
threshold for the onset of persistent granular flow
\citep{charru04:_erosion,paphitis05:_armouring}.

Even when the flow is laminar, the rough surface of the granular bed
yields a fluctuating local fluid stress, albeit deterministically
related to the time dependent realization of the surface packing.
Statistical methods are therefore necessary to completely characterise
the onset of granular flow.  Because we expect the surface packing to
have a transient component, so will the granular flux.  However, we
expect there to be a true threshold forcing below which there exist
surface packings such that the associated surface stress is
insufficient to dislodge any grains.  Under steady pre-threshold
forcing conditions, one such surface packing will eventually be
realized and the granular flow will cease.  We are interested in
predicting the value of the threshold fluid forcing, the transient
granular flux in pre-threshold conditions, and the steady state
granular flux above the threshold.  We would also like to establish
whether a single characteristic of the fluid forcing (such as the
surface stress) is sufficient to predict the steady state granular
flux.

Here we report our initial investigation of these questions in a cell
partially filled with transparent spherical glass beads driven by an
index-matched fluid.  The small size of the cell and the grains and
the high viscosity of the fluid ensures that the particle Reynolds
number is always less than ten.  Adding small fluorescing tracers and
illuminating the cell with a laser sheet allows us to directly image
all grains in a vertical slice through the system.  We can therefore
simultaneously measure the fluid and granular fluxes, the dimensions
of the fluid filled region, and the granular packing.  This allows us
to test theoretical predictions of the threshold condition and the
functional dependence of the granular flux on fluid driving.  Our
study may be considered complementary to that of
\cite{goharzadeh05:_trans}, who examined the fluid flow near a
granular bed using a similar setup.

\section{Experimental technique}
\label{sec:exper-techn}

\begin{figure}
  \centering
  \includegraphics[width=\linewidth]{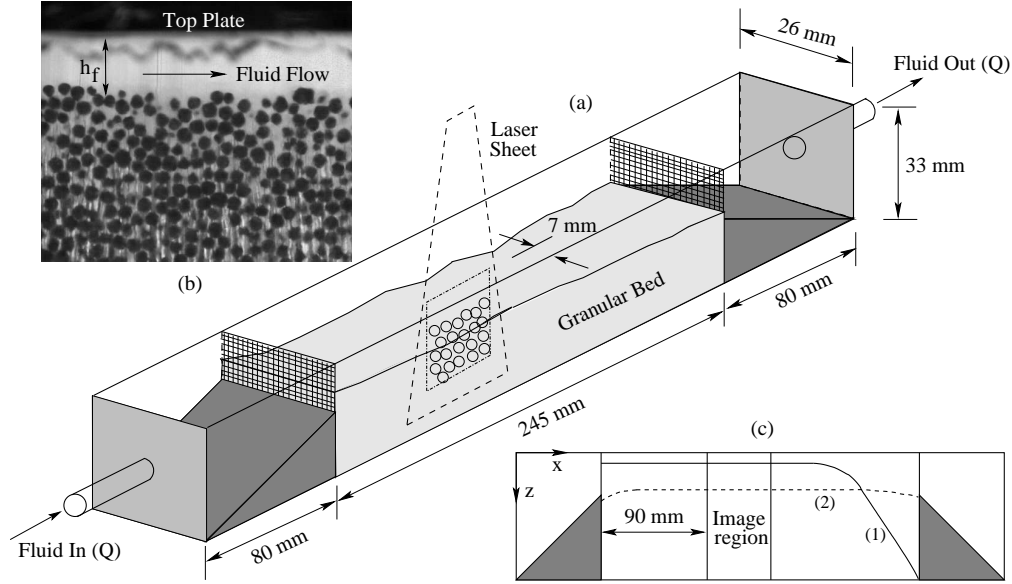}
  \caption{(a) Schematic diagram of the experimental apparatus. (b)
    Sample image of the refractive index matched beads in a plane away
    from the side walls.  (c) Side view of the experimental apparatus.
    Solid line (1) is the initial profile of the granular bed after
    the preparation steps (i)--(iii) are carried out.  Dashed line (2)
    is a sketch of typical profile of the granular bed after the
    cessation of erosion, the profile is perfectly flat except near
    the ends.  The image region is located about 90 mm (130d) from the upstream end.  \label{fig1}}
\end{figure}

As shown in figure~\ref{fig1}, an enclosed cell with a rectangular
cross-section is partially filled with glass beads with diameter $d =
0.70 \pm 0.05$ mm, density $\rho_g = 2.5$ g/cm$^3$ and index of
refraction of $n = 1.54$.  A peristaltic pump is used to generate a
constant fluid flux $Q$ through the cell.  The fluid is manufactured
to have the same refractive index as the glass beads and a dye is
added which fluoresces with a peak intensity at a wavelength of 565 nm
when illuminated by light with a wavelength of 532 nm
\citep{siavoshi06}.  The hydrocarbon based fluid, manufactured by
Cargille, Inc, has density $\rho_f = 1.026$ g/cm$^3$ and kinematic
viscosity $\nu = 0.243$ cm$^2$/sec.

The granular bed is prepared using the following three steps.  (i) we
tilt the cell vertically with the fluid input end at the bottom, allow
the grains to come to rest and (ii) slowly set it down horizontally.
This procedure fills the cell top leaving an empty buffer space near
the exit mesh (see figure~\ref{fig1}c).  Step (iii) of bed preparation
involves applying a small fluid flux to create a small (3--4 grain
diameters) uniform gap.  The three steps above are repeated to create
the initial condition for all runs.  We scanned the laser sheet across
the initial bed surface and verified that it was flat in the
observation window of width to within half a bead diameter.  After the
bed is prepared, the fluid flux is instantaneously raised to some
value $Q$ and kept constant for the duration of each run.  After traversing the pile's surface, eroded grains fall out of the way into the empty buffer space between the pile and the downstream mesh. Eventually, all granular
motion ceases.  In such a rest state the profile is perfectly flat
except in regions of order 2 cm near the inlet and outlet meshes.  We
therefore suppose, and argue further below, that in the flat region,
which comprises at least a 20 cm portion of the cell, the fluid flow
profile is steady, laminar and independent of the downstream distance.
We further checked that systematically varying the initialization fluid flux to create different initial bed heights did not change the final bed height. We observed that the final height reached was the same to within a grain diameter, provided the initial bed height was greater.

The range of fluid fluxes $Q$ for which the rest state is reached is a
function of the cell filling fraction.  If the cell filling fraction
is high, the range of $Q$ is rather small.  However, if the filling
fraction is small, the quiescent bed does not have a flat region when
the rest state is reached.  We therefore set the uniform bed level at
roughly 22 mm.

A laser and cylindrical lens system, placed above the cell,
illuminates a vertical slice through the cell away from the sidewalls.
Because the beads do not contain the dye, they appear dark against a
bright background fluid.  A high speed digital camera with a
resolution of 512$\times$480 pixels records a sequence of images at a
rate of 30--60 frames per second.  A typical image of a vertical slice
through the bed is shown in figure~\ref{fig1}b.  The image is
truncated at a depth of about 15 mm (measured from the top plate)
below which the particles never move.  The apparent size of a bead
depends on the distance of its centre from the illuminated plane.  We
are able to identify all beads whose centres are within approximately
$0.3d$ from the illuminated plane.  The height of the granular bed in
every image is measured using an edge detection algorithm.  We take
the height of the granular bed to be the inflection point in a depth
profile of the pixel intensity averaged over the image width.  This
technique yields the mean height of the bed in the observation window.
Furthermore, the mean height is averaged over 5 experimental runs to reduce statistical fluctuations, and remove any variability in preparing the initial pile.

We verified that the bed height decreases monotonically away from the
side walls up to a distance of about 3.5 mm beyond which it remains
constant to within the roughness scale of the granular bed.  For all
subsequent measurements we therefore positioned the laser sheet at a
distance 7 mm from the side walls.  The imaged region is about 16 mm
wide, and is located 90 mm downstream from the inlet mesh (see
figure~\ref{fig1}c). Comparing the height data for experimental runs repeated under same preparation conditions, we observed the standard deviations to be less that one particle diameter.

The inlet and outlet ramps ensure a smooth flow of the fluid onto the
granular bed.  After a process of trial and error, the height of the
ramps (22 mm) was chosen to be slightly lower than the height of the
granular bed when the grains fill the cell uniformly forming a flat
surface.  The leading edge of the bed tends to align with the top of
the inlet ramp.  If the bed is eroded below the upper edge of the
inlet ramp, an undesirable turbulent region can form behind it.
Neglecting end effects is theoretically justified because we find a
relationship between the grain flux and the \textit{local} viscous
shear stress in a region away from both ramps.  In addition, we find
that the final rest height of the bed away from the ramps depends only
on fluid flux $Q$ and not on the bed preparation or on the geometry of
the ramps.  Therefore, we maintain that the ramps only affect the
granular bed in their vicinity and do not influence the bed height
further than a certain distance $\ell_b$ downstream.  To estimate
$\ell_b$, we identify the length scale at which the width $\delta(x)$
of the Blasius boundary layer equals half the fluid depth $h_f$.  The
characteristic flow velocity is $U \sim Q/W h_f$, where $W$ is the
width of the cell.  Therefore
\begin{equation}
  \label{eq:3}
  \ell_b \sim \frac{h_f^2 U}{\nu} \sim  \frac{Q h_f}{4W\nu}.
\end{equation}
For the largest values of the flux $Q$ and the fluid depth $h_f$ in
our experiment, the length $\ell_b \approx 100d$.  Our use of the
laminar flow profile in the window of observation located $130d$
downstream from the inlet ramp is therefore justified.

\begin{figure}
  \centering
  \includegraphics[width=6cm]{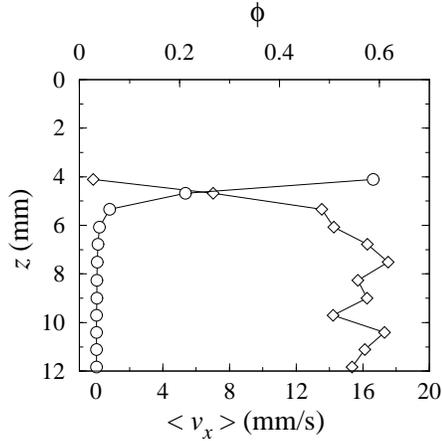}
  \caption{The horizontal granular velocity $\langle v_x \rangle$
    (bottom axis, circles), computed in the regime when grains are
    rolling on the surface of a quiescent pile, averaged over the
    observation window and a one second time interval.  The granular
    volume fraction $\phi$ is plotted with diamonds (top axis).  The
    vertical axis denotes the depth $z$ measured from the cell's top
    plate; $z=0$ corresponds to the top plate; the bed's surface is
    located where $\phi$ vanishes.  The total granular flux is
    proportional to the $z$ integral of the product of $\langle v_x
    \rangle \phi$.  \label{fig2}}
\end{figure}

To compute the vertical profile of the particle velocity and packing
fraction we measure the position of every particle in the image to
within $0.1d$ using a centroid technique
\citep{siavoshi06,tsai03:_shear_cryst}.  The velocity $v_x$ of
individual particles is then determined by tracking the particles over
two successive images.  The image region is divided into horizontal
bins of height $1d$.  The volume fraction $\phi$ is determined by
dividing the number of particles in each bin by the volume of the bin
($1d\times 0.6d\times 22d$).  The depth dependence of the bin-averaged
and time averaged (over approximately one second) velocity $\langle
v_x \rangle$ and volume fraction $\phi$ for one particular flow rate
is shown in figure~\ref{fig2}.  Above a certain depth no grains are
detected in the bin which yields a null packing fraction.  Given the
depth profiles of the horizontal velocity and volume fraction, the
grain flux $q_g$ (averaged over the observation window) is obtained by
computing the integral $\int \langle v_x \rangle \phi \ dz$.  Every
reported flux value represents the average over a time interval of one
second within which the variation in the flux values is small.  All
the experiments were repeated five times and the reported values are
the averages over these five runs.

Before we launch into the detailed analysis of the data, let us
qualitatively describe the phenomenon. Immediately after the fluid
flow is switched on, the granular bed experiences a brief period (a
few seconds) of rapid shear flow during which several grain layers are
moving.  The fluid gap increases rapidly during the shear flow regime.
Subsequently, the granular flux is conveyed via ``bed-load'', i.e.\
individual grains rolling on the surface of an apparently quiescent
bed.  After several minutes the granular flux eventually vanishes.
Because the erosion is gradual, the fluid gap increases gradually.
Thus, the bead shear stress decreases gradually and the bed (away from
the ramps) must approach a trivial flat state in which the fluid
exerts a threshold shear stress on the bed.

\section{Rest height of the granular bed}
\label{sec:rest-height}

\begin{figure}
  \centering
  \includegraphics[width=0.95\linewidth]{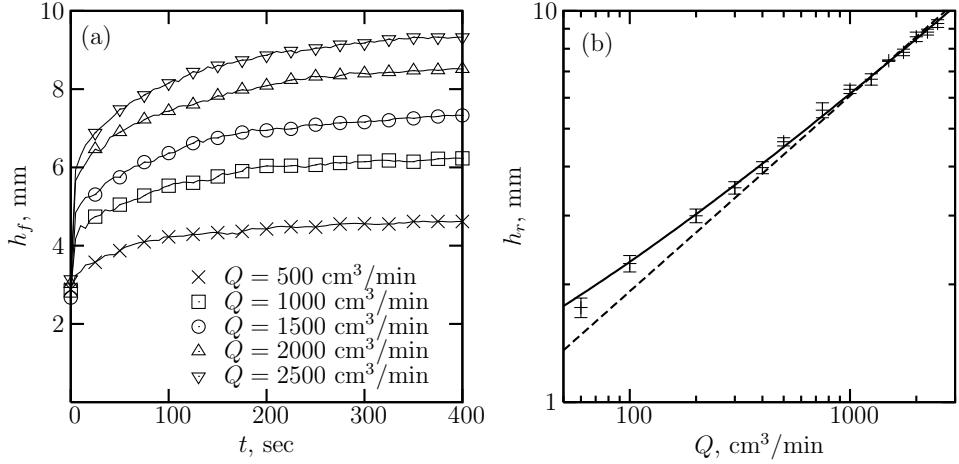}
  \caption{(a) The average depth $h_f$ of the fluid layer above the
    granular bed.  Solid lines are a guide to the eye.  (b) The rest
    depth $h_r$ of the fluid layer measured after cessation of
    granular flow as a function of fluid flux $Q$.  Dashed line is a
    fit to equation (\ref{eq:2}).  Solid line is the same fit with
    $h_r$ offset by roughly 0.4 mm.  For reference, the fluid depth
    above the entrance ramp is 11mm.  \label{fig3} }
\end{figure}

Figure \ref{fig3}a shows the average depth of the fluid gap as a
function of time for a constant fluid flux $Q$ over the bed.  As the
grains are eroded from the bed's surface and deposited into the buffer
space, the fluid gap grows, and the mean fluid flow speed and
therefore the shear stress on the bed decreases.  The erosion
eventually ceases and the fluid gap approaches a constant rest depth
$h_r$ which decreases with the applied fluid flux $Q$.

The gradually increasing fluid gap and thus gradually decreasing bed
stress leads to the cessation of erosion precisely at the threshold
driving.  The variation of the rest depth $h_r$ with the applied fluid
flux $Q$ allows us to examine the validity of the Shields threshold
criterion.  We estimate the bed shear stress $\sigma$ assuming
two-dimensional laminar flow between infinite smooth parallel plates
separated by $h_f$.  This approximation is good when the cell width $W
\approx 26$ mm is much greater than the fluid gap depth $h_f$.  Using
the solution for laminar flow in a pipe of rectangular cross-section
(see for example \cite{cornish28:_rect_pipe}) we can compute the
correction to the estimated bed stress.  Neglecting the sidewalls
results in an underestimate of the stress by 8\% for the smallest
fluid gap and by 29\% for the largest fluid gap in our experiment.
However, neglecting the sidewalls has the advantage of a simple
analytic expression for the bed stress which greatly simplifies the
analysis.  The laminar flow assumption is good since the cell Reynolds
number for $Q = 1000$~cm$^3$/min is $\mathrm{Re} = v_\mathrm{mean}
h_f/\nu = Q/W\nu \approx 0.3$, where $v_\mathrm{mean}$ is the
horizontal fluid velocity averaged over the fluid gap.  The Shields
parameter $\theta$, defined as the stress $\sigma$ scaled by the
hydrostatic pressure difference $(\rho_g - \rho_f) g d$ across the
grain, is
\begin{equation}
  \label{eq:1}
  \theta = \frac{6Q \nu}{W \gamma g d h_f^2},
\end{equation}
where $\gamma \equiv \rho_g/\rho_f - 1 \approx 1.437$ is the density
contrast.  The Shields parameter measures the relative importance of
the destabilising hydrodynamic forces and the stabilising gravity.

Assuming that the rest depth $h_r$ of the fluid gap corresponds to
the Shields criterion $\theta = \theta_c$, we obtain
\begin{equation}
  \label{eq:2}
  h_r = d\left(\frac{Q}{Q_r}\right)^{1/2}, \quad \mathrm{with} \quad
  Q_r = \theta_c  \frac{W \gamma gd^3}{6\nu}.
\end{equation}
The fit to the measured rest depth (shown as a dashed line in
figure~\ref{fig3}b) is significantly improved if a positive constant
of approximately half a diameter $d$ is introduced to the right-hand
side of (\ref{eq:2}) (an even better fit is obtained if $h_r \sim
Q^{0.45}$).  The fit yields $Q_r = 15.4 \pm 0.3$ cm$^3$/min which
translates to the critical Shields parameter of $\theta_c = 0.30 \pm
0.01$.  This value is consistent with previously reported values, for
example, in \cite{miller77:_shields_review} or
\cite{buffington97:_systematic}.  For comparison, the Yalin parameter,
$\Xi \equiv (\gamma g d^3)^{1/2}/\nu$, a less widely used
dimensionless group which measures the relative importance of viscous
and gravitational forces without reference to the flow intensity, is
$\Xi \approx 0.03$ in our experiment.

The physical origin of the offset is not entirely clear to us at this
time.  The correction to the bed stress estimate due to the sidewalls
cannot be the source of the offset since in the limit of small gap
$h_f$ the sidewall effect vanishes.  A possible source of the offset
is the fluid boundary condition at a permeable wall which relates the
derivative of the fluid velocity to its magnitude at some penetration
length scale $\ell$.  Yet another possibility is that the offset could
be due to the way in which $h_f$ is measured.  We use the mean surface
height to represent the mean boundary stress.  In reality, however,
fluid flow past a rough permeable medium creates a distribution of bed
stresses and the estimate of its mean using a non-slip boundary
condition at a flat wall is a gross approximation.  The fact that the
error in the estimate of the mean stress translates to a height offset
that is only half a bead diameter is a pleasant surprise.

\section{Flow rule}
\label{sec:flow-rule}

\begin{figure}
  \centering
  \includegraphics[width=8cm]{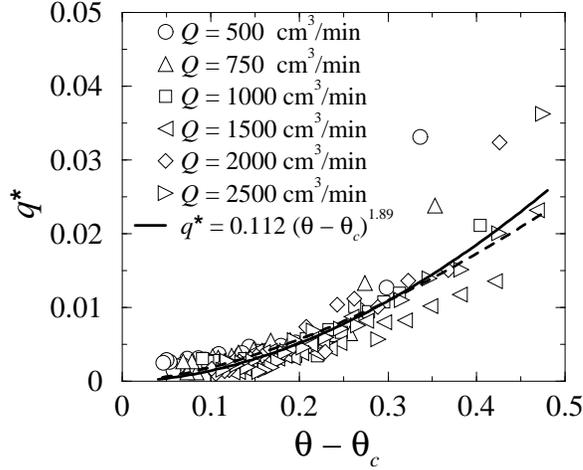}
  \caption{Einstein number $q^*$ vs.\ the excess $\theta - \theta_c$
    of the dimensionless shear stress.  Solid line is a power-law fit
    to the region $\theta - \theta_c < 0.3$.  Dashed line is a power
    law fit with $\lambda = 1.6$ fixed. \label{fig4} }
\end{figure}

Tracking all grains in a vertical slice as mentioned in
section~\ref{sec:exper-techn} allows us to directly measure the grain
flux $q_g$ and to attempt to relate it to the estimated mean bed
stress.  In general, other factors besides bed stress, such as the
flow history, could influence the grain flux.  Although an unambiguous
relationship between grain flux and bed stress is usually tacitly
assumed, here we test it directly.

Following convention \citep{einstein50} we use the Einstein number
$q^* = q_g/(\gamma g d^3)^{1/2}$, i.e.\ the non-dimensionalised grain
flux.  Figure~\ref{fig4} is a plot of $q^*$ vs.\ the estimated excess
bed stress $\theta - \theta_c$ computed using equation (\ref{eq:1})
with the value of $\theta_c = 0.3$ measured in
Section~\ref{sec:rest-height}.  Different symbols correspond to
different fluid fluxes $Q$.  Because the same bed stress occurs at
different fluid gap depths for different $Q$ and therefore entirely
different times in the history of the erosion process, data collapse
suggests that only the instantaneous bed stress (and not its history,
for example) is required to predict the instantaneous grain flux.
Therefore, upon changes (in time or space) of the bed stress the grain
flux quickly relaxes to its corresponding value.  We hypothesise from
dimensional considerations that the relaxation time scale is $d^2/\nu
\approx 0.02$ sec, which is small indeed.

The data in Figure~\ref{fig4} do not collapse for high stresses, i.e.\
early times, because small fluid gaps are associated with changes in
the mechanisms of granular transport.  At early times the granular
flux occurs primarily as bulk shear of the granular pile, whereas
later in the history of bed evolution the grain flux is conveyed
through isolated grains rolling over the otherwise immobile granular
bed.  The latter form of transport is known as the ``bed-load''
regime.  A power law fit to the data collapse yields
\begin{equation}
  \label{eq:4}
  q^* \approx A\,(\theta - \theta_c)^\lambda, \quad \mathrm{with}
  \quad \lambda = 1.89 \pm 0.25, \quad A = 0.11 \pm 0.03.
\end{equation}
Our measured exponent $\lambda$ is slightly higher than the widely
accepted value of 1.5 \citep{meyer-peter48}, and also above the value
of 1.6 found by \cite{wong06:_reanal_meyer_peter} in their reanalysis
of \cite{meyer-peter48}.  The pre-factor $A$, however, is more than an
order of magnitude smaller than that quoted by
\cite{wong06:_reanal_meyer_peter} for turbulent flow.  The large
discrepancy suggests that although the granular flux in the laminar
fluid flow regime is described by the same functional dependence on
the Shields parameter as in the turbulent regime, the pre-factor is
not.  The turbulent fluid appears to be significantly more efficient
at entrainment and transport of surface grains.  This is perhaps not
surprising since for the same mean bed stress, the turbulent bursts
provide a more effective dislodging process than the smoothly varying
bed stress in the laminar flow regime.

\section{Approach to rest height}
\label{sec:approach-rest-height}

The experimentally verified relationship between the bed stress and
the granular flux allows us to compute the entire history of the fluid
gap depth $h_f(x,t)$ ($x$ is the coordinate in the flow direction) as
it approaches the rest depth $h_r$.  The mean thickness of the flowing
grain layer is negligible compared to $h_f$ in the ``bed-load''
regime.  Therefore, we write the conservation of sediment as
\begin{equation}
  \label{eq:5}
  \frac{\partial h_f}{\partial t} = \frac{\partial q_g}{\partial x}.
\end{equation}
To study the approach of the profile to the rest height $h_r$, we
introduce $\xi = h_f/h_r$, scale all lengths by $d$ and times by
$\sqrt{d/\gamma g}$ and use the flow rule (\ref{eq:4}) which can be
expressed as $q^* = A\theta_c^\lambda (1/\xi^2 - 1)^\lambda$.
Eq.~(\ref{eq:5}) can therefore be written as
\begin{equation}
  \label{eq:6}
  \frac{\partial \xi}{\partial t} + c(\xi) \, \frac{\partial
    \xi}{\partial x} = 0, \quad c(\xi) = 2\lambda A \theta_c^\lambda
  \, \xi^{-1-2\lambda}(1 - \xi^2)^{\lambda - 1} \, \frac{d}{h_r}.
\end{equation}
Given an initial condition $\xi(x,t_0) = \xi_0(x)$, the solution to
equation (\ref{eq:6}) (obtained by the method of characteristics)
determines $\xi(x,t)$ implicitly through $\xi(x,t) = \xi_0(x -
c(\xi(x,t)) \,t)$.

There are two issues which make the full solution of (\ref{eq:6})
difficult to use.  First, a possibility exists that the granular flux
includes a term proportional to the gradient of the bed height
resulting in a diffusive term in (\ref{eq:6}).  Our measurements of
the bed height are not sufficiently resolved to verify or reject such
a diffusive term.  Second, the domain of applicability of equation
(\ref{eq:6}) is limited in time and space.  In the beginning of the
experiment, granular transport is mainly conveyed via bulk shear and
thus does not obey (\ref{eq:4}).  Also, near the inlet ramp the bed
height may change rapidly and thus the approximate expression for the
shear stress (\ref{eq:1}) is invalid there.  Fixing the initial and
boundary conditions for equation (\ref{eq:6}) is therefore not
feasible.  Thus the full solution is not useful.

However, the long time asymptotic solution to (\ref{eq:6}) with a
small diffusive term may be insensitive to the initial and boundary
conditions.  Therefore, we seek a separable solution to (\ref{eq:6})
in the limit $\xi \rightarrow 1$ (i.e.\ nearly flat bed)
\begin{equation}
  \xi(x,t) = 1 - \left( \frac{x - x_0}{B\,(t - t_0)}
  \right)^{1/(\lambda-1)},
  \label{eq:7}
\end{equation}
where $x_0$ and $t_0$ are integration constants and $B = \lambda
A(2\theta_c)^\lambda d/h_r$.

\begin{figure}
  \centering
  \includegraphics[width=8cm]{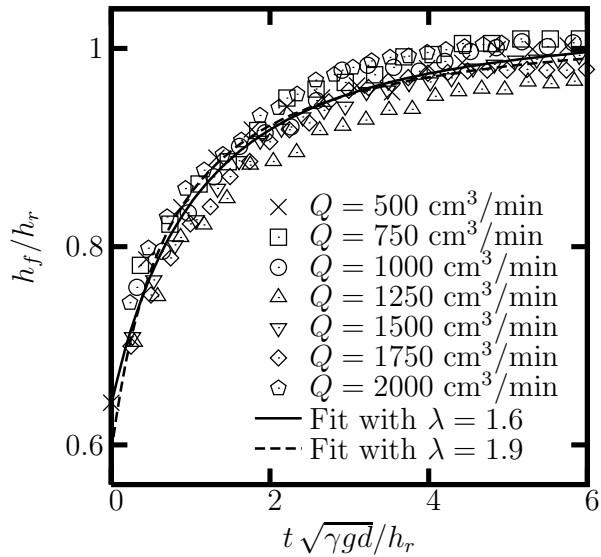}
  \caption{Fluid gap depth scaled by the rest depth $h_r$ vs.\ time
    scaled by $h_r/\sqrt{\gamma g d}$.  Solid line is a fit to a
    function that approaches unity as $(t - t_0)^{1/(1-\lambda)}$ with
    $\lambda = 1.6$ and $t_0 = -2.2 \, h_r/\sqrt{\gamma g d}$.  For
    comparison, we include a fit to a power law with $\lambda = 1.9$
    (dashed line).  \label{fig5}}
\end{figure}

Because the height of the bed changes by less than a diameter along the
cell, an experimental test of the spatial variation in
Eq.~(\ref{eq:7}) is impractical in our setup.  We therefore focus on
the approach of the fluid depth averaged over the observation window
to the rest depth $h_r$ (cf.\ figure~\ref{fig3}a).  Equation
(\ref{eq:7}) implies that once the fluid gap depth $h_f$ is scaled by
rest depth $h_r$ and time is scaled by $h_r/\sqrt{\gamma gd}$, the
data should collapse onto a master curve which approaches unity as the
power law $(t - t_0)^{1/(1 - \lambda)}$.  Figure \ref{fig5} shows the
data collapse.  The power law fit to the resulting data cloud yields
the offset time $t_0 \approx -2.2\, h_r/\sqrt{\gamma g d}$ and the
exponent $\lambda = 1.6 \pm 0.1$.  The predicted data collapse
suggests that the assumed functional relationship between the
instantaneous bed stress and granular flux is indeed correct.  It is
difficult to interpret the value of $t_0$, since the asymptotic
solution~(\ref{eq:7}) only applies at long times.  The fitted value
for $\lambda$ falls just outside the 1$\sigma$ interval of the value
quoted in (\ref{eq:4}), but is consistent with the value found by
\cite{wong06:_reanal_meyer_peter}.

\section{Discussion}
\label{sec:discussion}

In summary, we have revisited the well-studied problem of a sand bed
driven by an overlying viscous fluid.  We focused on precise
measurements in the laminar flow regime where little data is
available.  Our flow geometry, in which the depth of the fluid gap
gradually increases as the grains are eroded, results in the cessation
of granular flow and thus allows for an unambiguous definition of the
threshold condition.  Therefore, by varying the fluid flux $Q$ through
the cell, we have verified the hypothesis that the boundary viscous
shear stress is sufficient to predict cessation of granular flow over
the surface of a horizontal granular surface.  The directly measured
value of the critical Shields parameter $\theta_c = 0.30 \pm 0.01$ and
the Yalin parameter $\Xi = 0.03$ is consistent with the extrapolation
of the previously reported values of $\theta_c$ to this value of
$\Xi$.  The advantage of approaching the threshold from above is that
there is no ambiguity or subjectivity in its definition.  On the other
hand, a relatively rapid approach to the quiescent state in our setup
precludes a thorough study of the ``armouring'' phenomenon.  Finally,
we remark that the threshold for the cessation of granular flow may
differ from that for the onset of granular flow.  This effect is
difficult to quantify in general since the onset of granular flow is
marked by transient flow and steady driving is impossible in our cell
since any erosion leads to a decrease in driving.

We directly verified that the boundary shear stress determines not
only the onset of granular flow, but also the granular flux over a
roughly horizontal bed in non-steady driving conditions.  This is an
important result since it implies that the granular flux quickly
adjusts to changes in the driving.  We have established the power law
relationship between the grain flux and the excess boundary shear
stress via two independent measurements.  We measure the granular flux
directly by particle tracking and, alternatively, we analyse the
approach of the bed height to the quiescent state.  The average of the
two independently measured exponents is $\lambda = 1.75 \pm 0.25$.
The pre-factor in the power law relationship between the Einstein
number and the excess Shields stress is more than an order of
magnitude smaller than that in the widely used ``bed-load'' granular
flux rule \citep{wong06:_reanal_meyer_peter}, indicating perhaps a
different mechanism for the granular transport in the laminar fluid
flow regime.

We have measured the parameters in the granular flux rule (\ref{eq:4})
not only for a particular range of Reynolds numbers, but also for a
particular type of grains---smooth spheroids.  How the parameters in
(\ref{eq:4}) depend on the grain properties such as shape, friction
constant and roughness is an interesting and open question.

Lastly, we note that since the threshold condition in our experiment
is reached gradually, our problem is similar to the relaxation of a
dry granular pile expressed in terms of the angle of repose.
Moreover, the concept of a Shields threshold, typically defined via
the onset of granular flow due to gradually increased fluid driving,
may be analogous to the notion that dry granular piles can be
characterised by a maximum angle of stability.

\begin{acknowledgments}
  The work was funded by the Department of Energy grants
  DE-FG0202ER15367 (Clark), DE-FG0299ER15004 (MIT), and the National
  Science Foundation grant number CTS-0334587 (Clark).
\end{acknowledgments}

\bibliographystyle{jfm2}
\bibliography{/home/leapfrog/Manuscripts/all}

\end{document}